\documentclass[aps,prr,showpacs,superscriptaddress,twocolumn,raggedfooter,raggedbottom]{revtex4-1}   

\usepackage{amssymb}
\usepackage{amsmath}
\usepackage{amssymb}
\usepackage{graphicx}
\usepackage{subfigure}
\usepackage{color}
\usepackage{mathrsfs} 
\usepackage{float}

\usepackage{soul}

\usepackage[usenames,dvipsnames]{xcolor}
\usepackage[colorlinks=true,citecolor=Blue,linkcolor=RubineRed,urlcolor=Blue]{hyperref}

\newcommand{\panos}[1]{\todo[inline,caption={},color=yellow!40]{#1}}
\usepackage{xcolor}
\renewcommand{\panos}[1]{{\color{black} #1}}
\begin{document}

\author{M. Beau}
\affiliation{Department of Physics, University of Massachusetts, Boston, Massachusetts 02125, USA}
\affiliation{Dublin Institute for Advanced Studies, School of Theoretical Physics, 10 Burlington Road, Dublin 4, Ireland}

\author{A. del Campo}

\affiliation{Department  of  Physics  and  Materials  Science,  University  of  Luxembourg,  L-1511  Luxembourg, Luxembourg}

\affiliation{Donostia International Physics Center, E-20018 San Sebasti{\'a}n, Spain}

\affiliation{Department of Physics, University of Massachusetts, Boston, Massachusetts 02125, USA}

\author{D. J. Frantzeskakis}
\affiliation{Department of Physics, National and Kapodistrian University of Athens,
Panepistimiopolis, Zografos, Athens 15784, Greece}

\author{T. P. Horikis}
\affiliation{Department of Mathematics, University of Ioannina, Ioannina 45110, Greece}

\author{P. G. Kevrekidis}
\affiliation{Department of Mathematics and Statistics, University of Massachusetts,
Amherst, MA 01003-4515, {USA}}

\title{Dark solitons in a trapped gas of long-range interacting bosons}

\begin{abstract}
We consider the interplay of repulsive
short-range and same-sign long-range interactions in the dynamics of dark solitons, as 
prototypical coherent nonlinear excitations in a trapped 1D Bose gas. 
First, the form of the ground state is examined, and then 
both the existence of the solitary waves 
and their stability properties are explored, and corroborated by direct numerical simulations. We 
find that single- and multiple-dark-soliton states can exist and are generically robust in the 
presence of long-range interactions. We analyze the modes of vibration of such excitations and find 
that their respective frequencies are significantly upshifted as the strength of the long-range 
interactions is increased. Indeed, 
we find that a prefactor of the long-range interactions considered
comparable to the trap strength may upshift the dark soliton
oscillation frequency by {\it an order of magnitude}, in comparison
to the well established one of $\Omega/\sqrt{2}$ in a trap of frequency $\Omega$.
\end{abstract}

\maketitle

\section{Introduction}

A paradigmatic model of one-dimensional bosons subject to contact interactions is known as the  Lieb-Liniger model (LL) \cite{LL63,Lieb63}. As an exactly-solvable model 
exhibiting scattering without diffraction \cite{Sutherland04}, it plays a crucial role in 
mathematical physics \cite{KBI97,Takahashi99,Gaudin14}. At the same time, it accurately describes  ultracold atomic clouds tightly confined in waveguides when interatomic scattering is dominated by the s-wave contribution \cite{Olshanii98,Cazalilla11}. 

Recently, it has been shown that a variant of the LL model admits an exact solution in the presence of a harmonic trap when the interparticle contact interactions are supplemented with a long-range term \cite{Beau20,delcampo20}. When the contact interactions are attractive, 
the long-range term is equivalent to a one-dimensional (1D) attractive gravitational potential.  By contrast, for repulsive contact interactions, the long-range term is equivalent to a 1D  repulsive Coulomb potential. The resulting long-range Lieb-Liniger (LRLL) model has intriguing connections with other physical models. Its ground-state wavefunction shares the structure of  Laughlin liquids of relevance to the fractional quantum Hall effect \cite{Lieb2018}. It also describes a 1D version of the non-relativistic Newtonian gravitational Schr\"odinger equation used in the modeling of dark matter as a self-gravitating Bose-Einstein condensate \cite{Broadhurst14}. 
In this context, soliton solutions are used to describe so-called ghostly galaxies, large and barely visible low-density galaxies, such as the dark-matter dominated Antlia II \cite{Broadhurst20}. 

The LRLL model is part of a larger class of solvable models that can be obtained as deformations of parent Hamiltonians by embedding them in a confining potential~\cite{delcampo20,Beau21}.  Such deformations are analogous to those known in the nonlinear-Schr\"odinger (NLS) equation 
\cite{Kundu09}. However, at the many-particle level, it is crucial that the embedded quantum state has a  Jastrow form, e.g., with a wavefunction expressed as a pairwise product of a correlation 
function~\cite{delcampo20,Beau21} over each pair of particles. The conventional LL model in free space with attractive interactions is solvable by Bethe ansatz and admits so-called string solutions with complex Bethe roots  \cite{Takahashi99}. In the center of mass frame, the lowest energy state was found by McGuire and describes a quantum bright soliton, a cluster of particles sharply localized in space  \cite{McGuire64}. Importantly, the McGuire bright soliton solution is given by a Jastrow form, making its embedding possible in a harmonic trap at the cost of supplementing the Hamiltonian with a two-body pairwise long-range interaction term. As a result, the trapped McGuire soliton is the ground state of the LRLL model in
the case of attractive interactions~\cite{Beau20}. 

When the many-particle wavefunction of a quantum state is not of Jastrow form,  embedding in a harmonic trap results in a parent Hamiltonian with many-body momentum-dependent interactions,  which need not be pairwise \cite{delcampo20}, and are less straightforward to justify on physical grounds.  This observation potentially precludes the investigation of dark solitons 
(namely density depressions, denoting the localized absence
of particles in space, accompanied by a phase jump across their density minimum) in the LRLL model.   Building on early results \cite{Kulish76,Ishikawa80,Tsuzuki71}, the investigation of many-body  quantum soliton wavefunctions for repulsive interactions in the absence of a trap has led to the  identification of a series of soliton-like quantum states \cite{Deguchi1,Deguchi2,GW00}. Yet, such  states lack the simple Jastrow structure required for their embedding in a trap to require solely  momentum-independent pairwise interactions.

This state of affairs is the starting point for our work. Can bosons with long-range interactions support dark 
soliton solutions in the mean-field regime? The LRLL mean-field limit was presented in 
Ref.~\cite{Beau20} and is described by a 1D NLS equation with a nonlocal nonlinearity. In the homogeneous space, it is known that the defocusing NLS model associated with a weakly-nonlocal repulsive interaction admits dark soliton solutions \cite{djf2}. In the case of the LRLL model as well as in its mean-field limit, the strength of the spatially-inhomogeneous harmonic confinement and the nonlocal nonlinearity are interrelated. 
This motivates our quest for dark soliton solutions in  a nontrivial inhomogeneous model of a trapped gas of long-range interacting bosons. 
 Specifically, we focus on
a NLS with local repulsive interactions and a nonlocal
long-range contribution of the same sign. This model is
inspired by the inhomogeneous NLS associated with the
mean-field theory of the LRLL, but there local and nonlocal interactions have opposite character, making the present
extension a nontrivial one.
We illustrate herein that a systematic characterization
of the underlying ground state can be offered under the interplay
of short-range and long-range  interactions.
Equipped with that, we can theoretically analyze the motion of the dark 
soliton on top of this background (and associated effective 
potential), by suitably adapting the methodology
of~\cite{konotop} to account for the presence of long-range 
terms.  We find that turning on even weak long-range
interactions has a drastic impact on the oscillation frequency
of the dark soliton in comparison to the frequency of the
confining parabolic potential. Upon extending these ideas to
multiple solitons,
we summarize our findings and present some directions
for future study.

\section{Analytical and Numerical Setup}
The regimes of degeneracy of a 1D Bose gas with contact interactions are well known since the seminal work by Petrov et al. \cite{Petrov00}. An analogous study for the recently-introduced LRLL model has not been yet performed. While the strength of the contact and long range interactions is characterized by a single common parameter, it is not possible to extrapolate the results from the case with only contact interactions to the LRLL model. In particular, the LRLL exhibits novel phases which are
absent in the conventional LL model. For instance, it can behave as an  incompressible Laughlin-like fluid with flat density or like a Wigner crystal~\cite{Beau20}. 
Chartering the phase diagram of the LRLL model remains an interesting prospect for further studies.

In this work, we take a different approach and focus on nonlinear physics inspired by the LRLL model. Specifically, motivated by the dynamical version of the mean-field model discussed in 
Ref.~\cite{Beau20}, we consider the following NLS equation (subscripts denote partial derivatives):
\begin{eqnarray}
i\hbar \Psi_t = &-&\frac{\hbar^2}{2m}\Psi_{xx} +g|\Psi|^2 \Psi + V(x)\Psi 
\nonumber \\ 
&+& ma \left( \int dx' |x-x'| |\Psi(x',t)|^2 \right)\Psi.
\label{NLS0}
\end{eqnarray} 
Here, $\Psi(x,t)$ is the mean-field wavefunction describing a 1D boson gas, consisting 
of atoms of mass $m$, confined in the parabolic trapping potential $V(x)=(1/2)\omega^2 x^2$ 
of frequency $\omega$. The atoms are assumed to interact repulsively via the contact 
(local) interaction, with coupling strength $g=2\hbar^2/(ma_s)$ (where $a_s>0$ is the 1D 
scattering length), as well as via the long-range (nonlocal) interaction, characterized by the 
effective coupling constant $a$ (with dimensional units of acceleration); this long-range effect 
can be  
induced either by gravitational attraction or Coulomb repulsion \cite{Beau20}.  
Next, measuring time, length and density $|\Psi|^2$ in units of $\omega_0^{-1}$, 
$a_0=\sqrt{\hbar/(m\omega_0)}$ and $2a_0^2/a_s$, respectively (where the  
frequency $\omega_0$ is a free parameter ---see below), we express Eq.~(\ref{NLS0}) in 
the following dimensionless form:
\begin{subequations}
\begin{eqnarray}
i \Psi_t &=&-\frac{1}{2}  \Psi_{xx} +  |\Psi|^2 \Psi + [V(x)+U] \Psi,  
\label{eqn1} \\
U &=& \beta \left(\int dx' |x-x'| |\Psi(x')|^2 \right), 
\label{eqn1b}
\end{eqnarray}
\end{subequations}
where the parabolic trapping potential now reads by $V(x)=(1/2)\Omega^2 x^2$, with the normalized 
frequency $\Omega$ and the parameter $\beta$ characterizing the long-range effect being given by:
\begin{equation}
\Omega=\frac{\omega}{\omega_0}, \quad \beta = \frac{a a_s}{2a_0^2 \omega_0^2}. 
\label{param}
\end{equation}
The model under consideration, Eqs.~(\ref{eqn1})-(\ref{eqn1b}), involves two parameters:  
the normalized trap frequency $\Omega$ and the normalized long-range interactions' strength $\beta$. 
In the case of $\Omega=\beta=0$, the system~(\ref{eqn1})-(\ref{eqn1b}) reduces to the completely 
integrable defocusing NLS equation, which possesses dark soliton solutions~\cite{djf,siambook}. 
In our analysis below, we will investigate the combined effect of the trapping potential and the 
nonlocal interactions to the dark soliton dynamics. It is clear that the relative magnitude of 
the parameters $\Omega$ and $\beta$, which both depend on the (undefined so far) frequency 
$\omega_0$, leads to different regimes, where the magnitude of $\omega_0$ can accordingly be 
estimated. Specifically, using Eq.~(\ref{param}), it can be found that, e.g., in the regime 
$\Omega \sim \beta$, the frequency $\omega_0=O\left(a a_s/(a_0^2 \omega)\right)$.
%
It is also noticed that 
using the Green's function identity $\frac{d^2}{dx^2} |x-x'|=2 \delta(x-x')$ 
(where $\delta(x)$ is the Dirac delta function), Eq.~(\ref{eqn1b}) leads to:
\begin{equation}
U_{xx}=2\beta |\Psi|^2,
\label{U}
\end{equation}
and hence the full integro-differential equation can alternatively be treated as the system 
of Eqs.~(\ref{eqn1}) and (\ref{U}).

The time-independent version of Eq.~(\ref{eqn1}) can be obtained upon using the standard ansatz  $\Psi(x,t)=\exp(-i \mu t) u(x)$, where $\mu$ is the chemical potential. In this way,   
we obtain the corresponding steady state problem for the function $u(x)$ in the form:
\begin{subequations}
\begin{eqnarray}
\mu u &=& -\frac{1}{2}  u_{xx} +  |u|^2 u + [V(x)+U] u, 
\label{adc2} \\
U &=& \beta \left(\int dx' |x-x'| |u(x')|^2 \right).
\label{adc2b}
\end{eqnarray}
\end{subequations}
Equation~(\ref{adc2}) is key to our analysis.
We focus herein on the case with 
$\beta \geq 0$,  
namely, we consider the interplay between repulsive short-range  and 
attractive  long-range  interactions. 
Our numerical computation starts from the local case {with} $\beta=0$ and
considers the \panos{approach to the} {Thomas-Fermi (TF)} limit of $\mu \gg \Omega$, \panos{in which the
role of the kinetic energy is becoming negligible}. In this limit, a well-defined
theory of dark solitons, analyzing their existence, stability, and dynamical properties, 
has been developed for quasi-1D BEC settings ---see, e.g. the reviews \cite{djf,siambook}. 
We obtain these dark solitons as (numerically) exact solutions up to a prescribed numerical tolerance, using a \panos{root finding algorithm (a Newton-Raphson scheme for the vector arising from the numerical
discretization of Eq.~(\ref{adc2})). An advantage of this method is
that it can be used in any regime, i.e., it is not restricted
to the Thomas-Fermi limit}. 

Subsequent consideration of the Bogolyubov-de Gennes (BdG) spectral analysis \cite{pethick,stringari}
 of the ground state and the solitons is then implemented using the perturbation ansatz:
\begin{eqnarray}
\!\!\!\!
\Psi(x,t)=e^{-i \mu t} \left[u(x) + \left(a(x) e^{i \omega t} + b^{\star}(x) 
e^{-i \omega^* t}\right) \right].
\label{bdg}
\end{eqnarray}
Here, $\omega=\omega_r + i \omega_i$ is the relevant eigenfrequency,
which when real indicates spectral stability (and oscillations with 
frequency $\omega_r$), while
if it has a nontrivial imaginary part $\omega_i \neq 0$, it indicates a dynamic instability with growth rate $\omega_i$. The pertinent eigenvector
$(a,b)^T$ corresponds to the eigendirection associated with the relevant
oscillation and/or growth. 
Once the solution of Eq.~(\ref{adc2}) is obtained, it is used as an input
in the eigenvalue solver resulting from the insertion of Eq.~(\ref{bdg})
into Eq.~(\ref{eqn1}), allowing us to assess the solution's spectral
features and its anticipated dynamical robustness.
Once the existence is obtained via Eq.~(\ref{adc2}) and the BdG stability
is characterized via Eq.~(\ref{bdg}), the solution is inserted in a 
dynamical integrator of Eq.~(\ref{eqn1}) [typically a fourth-order
Runge-Kutta in time, coupled with a second-order discretization in space]
to explore the dynamical properties of the waveform. 

\section{Ground State} 

To derive the ground state of the system characterized by a density $n(x)\geq 0$, we 
substitute $u=n^{1/2}$ into Eqs.~(\ref{adc2})-(\ref{adc2b}) and, 
using also Eq.~(\ref{U}), we obtain the following equations:
\begin{eqnarray}
&\frac{1}{2}n^{-1/2}(n^{1/2})_{xx} +\mu - n - V(x) - U = 0,
\label{h1} \\
&\panos{U = \beta \left(\int dx' |x-x'| n(x') \right)}.
\label{h2}
\end{eqnarray}
Below we are interested in finding the ground state in the {TF} limit, where 
the curvature term \panos{$(1/2) n^{-1/2}(n^{1/2})_{xx}$ [see Eq.~(\ref{h1})]} can be neglected 
\cite{pethick,stringari}. To be more specific, we seek a symmetric ground state, 
with $n(x)=n(-x)$, obeying the following normalization condition at $x=0$ (i.e., at the 
trap center):
\begin{equation}
n(0)+\beta \int dx' |x'| n(x') = \mu,
\label{nc}
\end{equation}
which stems from Eq.~(\ref{h1}). The above nonlinear boundary-value problem of Eqs.~(\ref{h1})-(\ref{h2}) for the 
ground-state density will be solved, following the same 
Newton-Raphson methodology as discussed above, for both the local 
($\beta=0$) and nonlocal ($\beta\ne 0$) cases.

\begin{figure}[tbp]
    \includegraphics[width=0.4\textwidth]{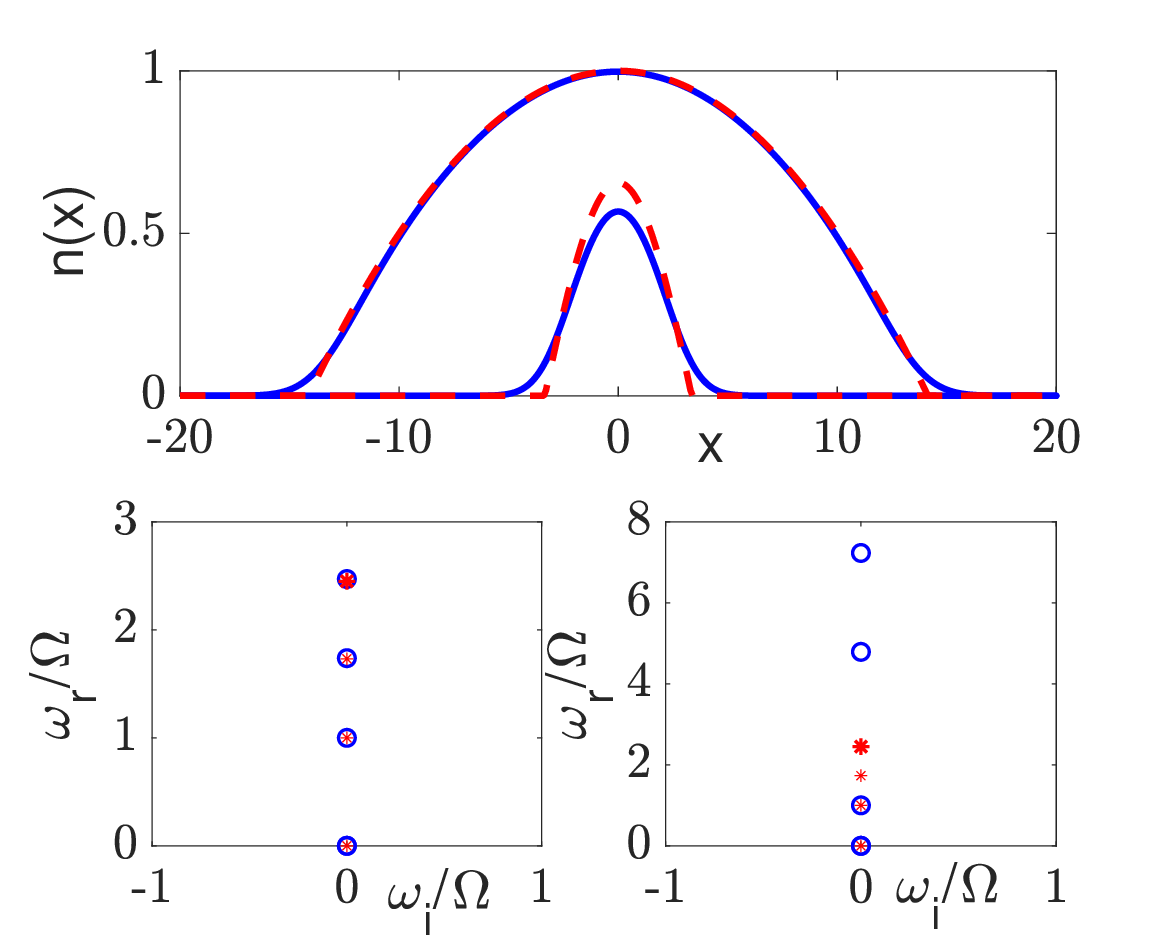}
    \includegraphics[width=0.4\textwidth]{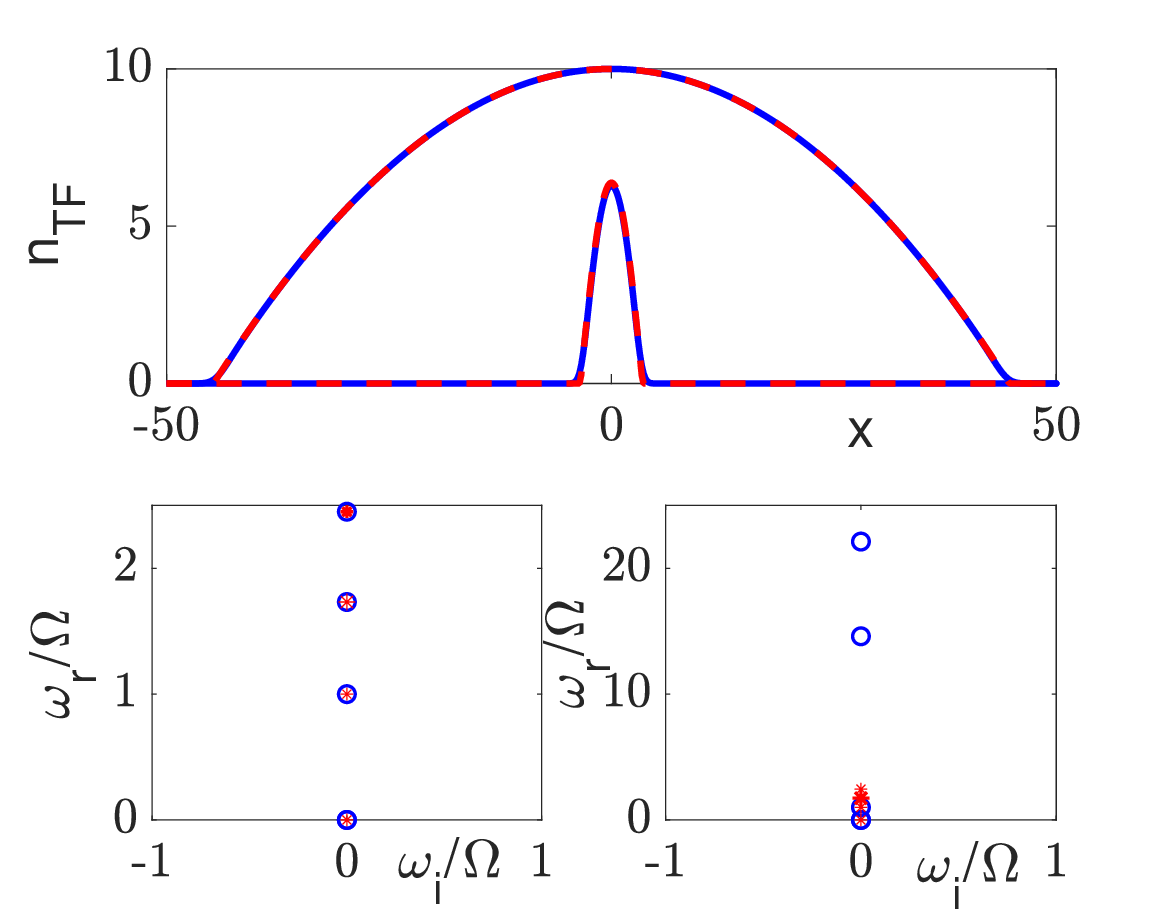}
    \caption{Ground state \panos{density} profile (top row) in the limit of
    $\mu=1 \gg \Omega=0.1$. The outer, inverted parabola
    profile corresponds to the (local) case of $\beta=0$~\cite{pethick,stringari}, \panos{as per Eq.~(\ref{ref_torture1})}. 
    The inner, smaller amplitude profile corresponds to the
    nonlocal case of $\beta=\Omega=0.1$. In both cases, 
    the solid blue line provides the numerical result,
    while the dashed red line corresponds to the analytical
    approximation. The second row presents the spectral
    plane ($\omega_i,\omega_r$) of the BdG eigenfrequencies
    $\omega=\omega_r + i \omega_i$
    for the case of $\beta=0$ (left) and $\beta=0.1$ (right). The
    numerically obtained $4$ lowest frequencies are shown {with} blue circles,
    while the analytical prediction of the TF limit for $\beta=0$
    i.e., $\omega/\Omega=\sqrt{m (m+1)/2}$~\cite{menotti,stringari,siambook,derosi} is shown
    with red stars. 
The absence of imaginary eigenfrequencies showcases
    the spectral stability of the corresponding configuration.
    The third and fourth row show the same features, but now
    for the case of $\mu=10 \gg \Omega=0.1$. 
    }
    \label{ex_gs}
\end{figure}

The limit of local interactions with  $\beta=0$ (i.e., $U=0$) 
is described by the defocusing NLS ($g>0$) and features 
a positive definite, nodeless ground state, with a {TF} density profile that 
can be found in the limit 
of $\mu \gg \Omega$ \cite{pethick,stringari}. 
Indeed, in this limit performing the standard approximation 
neglecting the second derivative term~\cite{pethick,stringari}, 
we obtain:
\begin{equation}
n_{\rm TF}(x)=\max\left\{ \left(\mu-\frac{1}{2}\Omega^2 x^2\right),0\right\}.
\label{ref_torture1}
\end{equation}
This expression  captures very accurately the core of the relevant distribution
and only ``falters'' at the low-density tails, 
where suitable asymptotic 
corrections can be devised~\cite{gallo}. The relevant stationary state for $\mu=1$ and 
the TF analytical approximation are shown as the larger (inverted parabola) profile in the top row
of Fig.~\ref{ex_gs}; see also the third row of the figure for the
case of $\mu=10$.

On the other hand, for the fully nonlocal case with $\beta\neq 0$, we may use a similar 
methodology and derive $n_{\rm TF}$. 
Indeed, we differentiate Eq.~(\ref{h1})  
twice with respect to $x$, and substitute $U_{xx}=2\beta n$ from Eq.~(\ref{h2}); then, 
in the TF limit, where the curvature term $[(1/2) n^{-1/2}(n^{1/2})_{xx}]_{xx}$ can be 
neglected, we obtain the following equation:
\begin{equation}
n_{xx} +2\beta n + \Omega^2= 0.
\label{eqn_tf3b}
\end{equation}
The symmetric solution of the above equation represents the TF density profile: 
\begin{eqnarray}
n_{\rm TF}(x)=A \cos \left( \sqrt{2 \beta} x \right) - \frac{\Omega^2}{2 \beta},
\label{eqn_tf1}
\end{eqnarray}
where $A$ is a constant. 
Naturally, and similarly to Eq.~(\ref{ref_torture1}), we note that
the density cannot become negative. Hence, the TF density   
consists of the {\it central lobe} of Eq.~(\ref{eqn_tf1}), while the
rest of the spatial domain is padded with a zero background.
In this case, the amplitude $A$ of the solution can be derived via the normalization 
condition~(\ref{nc}), namely by the following transcendental equation:
\begin{eqnarray}
&& \left(A -\frac{\Omega^2}{2 \beta}\right) 
\nonumber
\\
&+& 2 \beta \int_{0}^{L} 
x' \left[ A \cos \left( \sqrt{\frac{2 \beta}{g}} x' \right) 
- \frac{\Omega^2}{2 \beta} \right] dx' = \mu,
\label{eqn_tf2}
\end{eqnarray}
where $L=\sqrt{1/(2 \beta)} \cos^{-1}(\Omega^2/(2 \beta A))$ is the effective ``TF radius''.
We have solved this equation numerically for different
parameter values; e.g., for
$\beta=\Omega=0.1$ and $\mu=1$, we find $A=0.706$. 
This, then, enables
us to produce an approximate profile for the TF density which is also
compared with the corresponding numerical result in the top two rows of Fig.~\ref{ex_gs}.
The first row thereof presents the comparison of the relevant
density profiles, while the second row illustrates the collective  frequencies of the
BdG (stability) analysis for both cases, $\beta=0$ (left) and $\beta=0.1$ (right).
While the agreement is not as remarkable as in the local case 
(presumably due to the enhanced curvature of the solution, especially
near $x=0$), we still obtain a reasonable approximation of the corresponding ground state
profile. Indeed, this prompts one to think that, presumably, despite the
$\mu \gg \Omega$ setting, the TF limit has not been yet reached.
In light of that, we considered a far larger value of $\mu=10$, for
which repeating the calculation yields an analytical estimate
of $A=6.4371$ for $\beta=0.1$ (based on the
solution of Eq.~(\ref{eqn_tf2})). 
In that case, as can be seen in the third and fourth rows
of Fig.~\ref{ex_gs}, the analytical expression of Eq.~(\ref{eqn_tf1}) 
 captures very accurately the numerically obtained solution, not only
for $\beta=0$, but also for the nonlocal case of $\beta=0.1$.
It is also interesting to note that while the known frequencies
of the TF cloud in the absence of the nonlocal effect
$\omega/\Omega=\sqrt{m (m+1)/2}$ for positive integer $m$~\cite{menotti,stringari,siambook,derosi} are
precisely captured (see, e.g., the bottom left panel), there is a significant
upshift of the relevant frequencies (i.e., downshift of the period
of the respective modes) for $\beta=0.1$, as shown in the bottom
right panel of Fig.~\ref{ex_gs} \panos{both for $\mu=1$ and for $\mu=10$}. 

Armed with the above understanding of the ground
state of the system, we now turn our attention to the study of
dark soliton states.

\section{Single and Multiple Dark Solitons}

\begin{figure}[tbp]
    \includegraphics[width=0.4\textwidth]{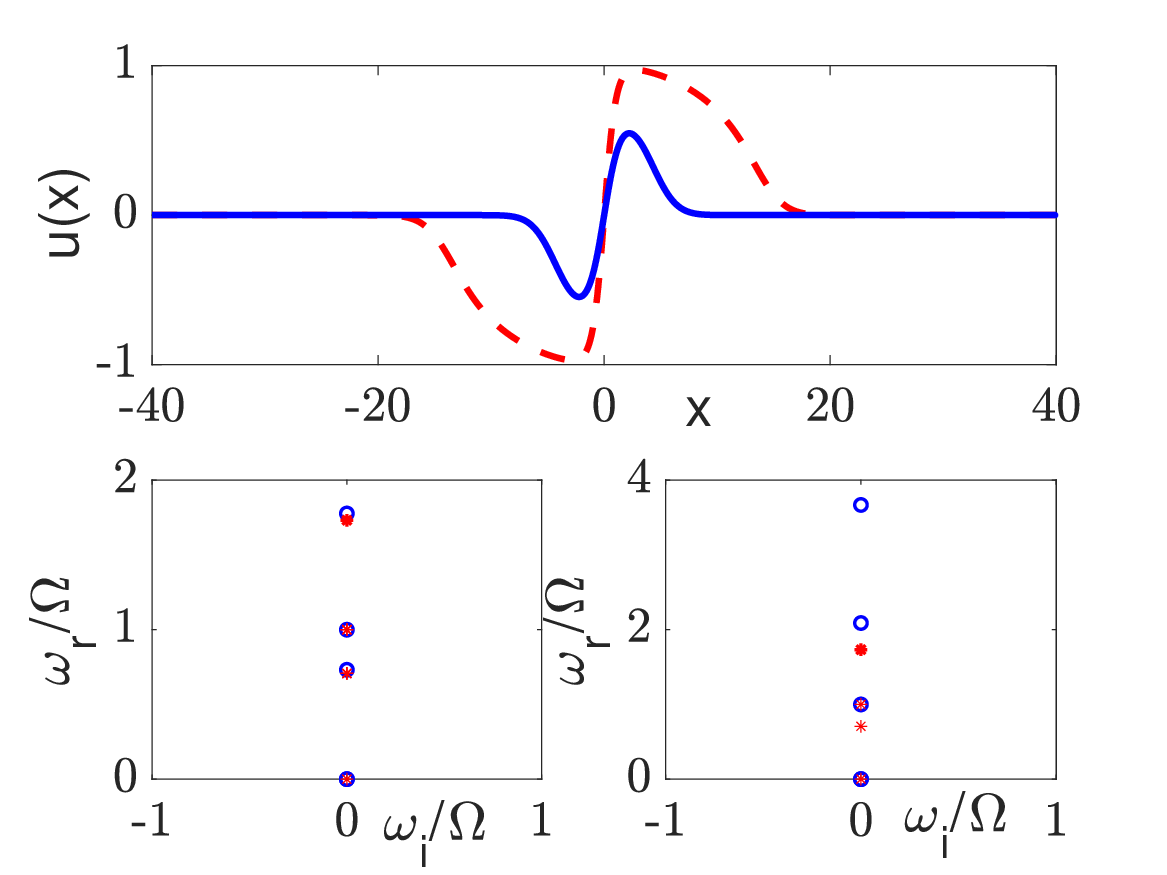}
    \includegraphics[width=0.4\textwidth]{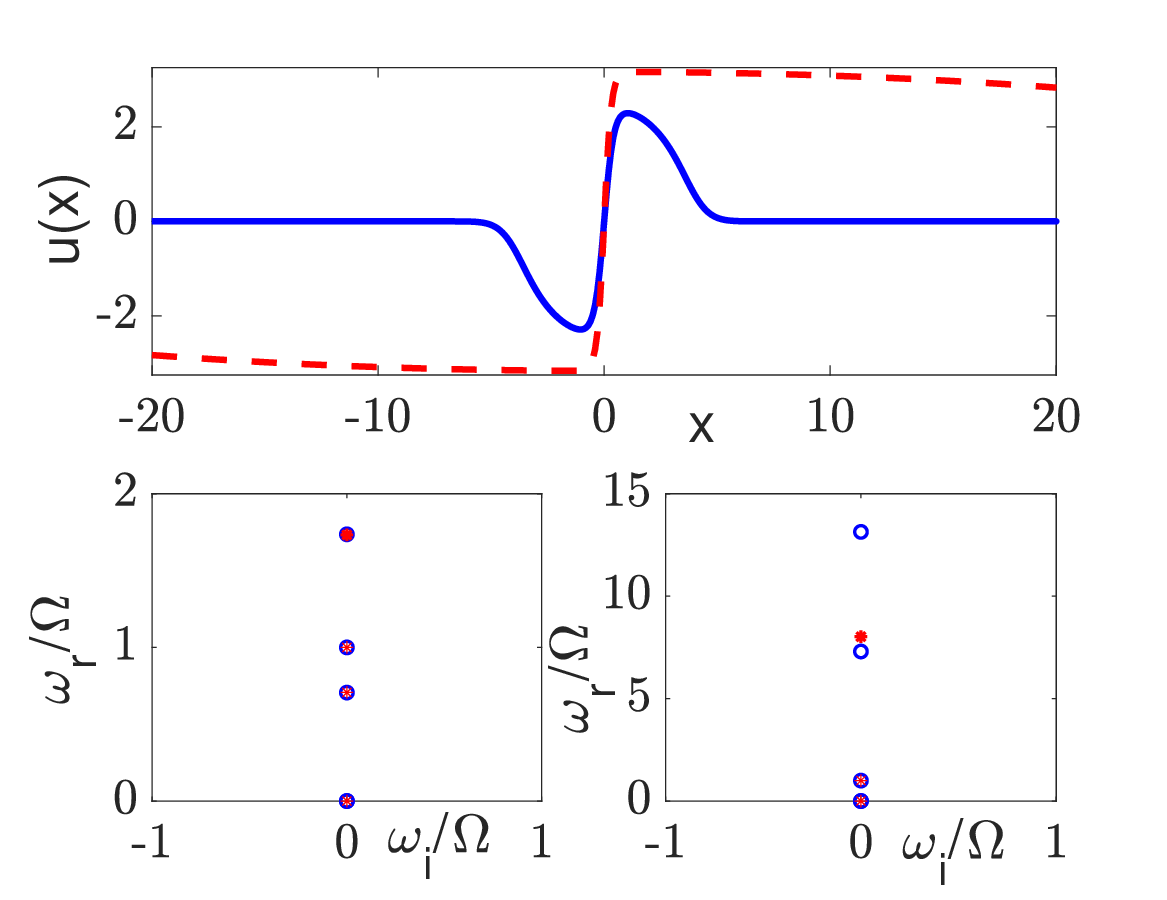}
    \caption{The top panel contains the exact stationary trapped
    dark soliton solution in the absence (i.e., $\beta=0$, dashed line)
    and in the presence ($\beta=\Omega$, solid line) of long-range interactions. \panos{The chemical potentials and trap parameters are
    directly analogous (in top and bottom panels) to those of Fig.~\ref{ex_gs}.}
    The second row panels show the BdG results (again the imaginary vs. the
    real part of the $4$ lowest eigenfrequencies) for $\beta=0$ (left panel)
    and $\beta=\Omega$ (right panel). The real nature of the eigenfrequencies
    indicates stability in both cases. The numerical results in both settings
    are indicated by blue circles. The red stars show in both cases 
    the analytical predictions
    in the TF limit for $\beta=0$ for comparison (see also text).
    The third row panel represents the solution for $\beta=0$ (dashed line) and
    $\beta=0.1$ (solid line) for the TF limit case of $\mu=10$. The bottom
    panels show the corresponding BdG eigenfrequencies 
   for $\beta=0$ (left) and $\beta=\Omega$ (right panel). Notice in the 
    bottom left panel 
    the coincidence of the numerical (blue circles)
    and analytically predicted (red stars--- see also text) frequencies.
    However, even in the nonlocal case of $\beta \neq 0$    
    of the bottom right
    panel, the symmetry modes at $\omega=0$ and $\omega=\Omega$ and the dark soliton
    vibrational mode (see text around Eq.~(\ref{kk_prl2})) are 
    theoretically captured.}
    \label{ex_stab_ds1}
\end{figure}

\begin{figure}[tbp]
    \includegraphics[width=0.4\textwidth]{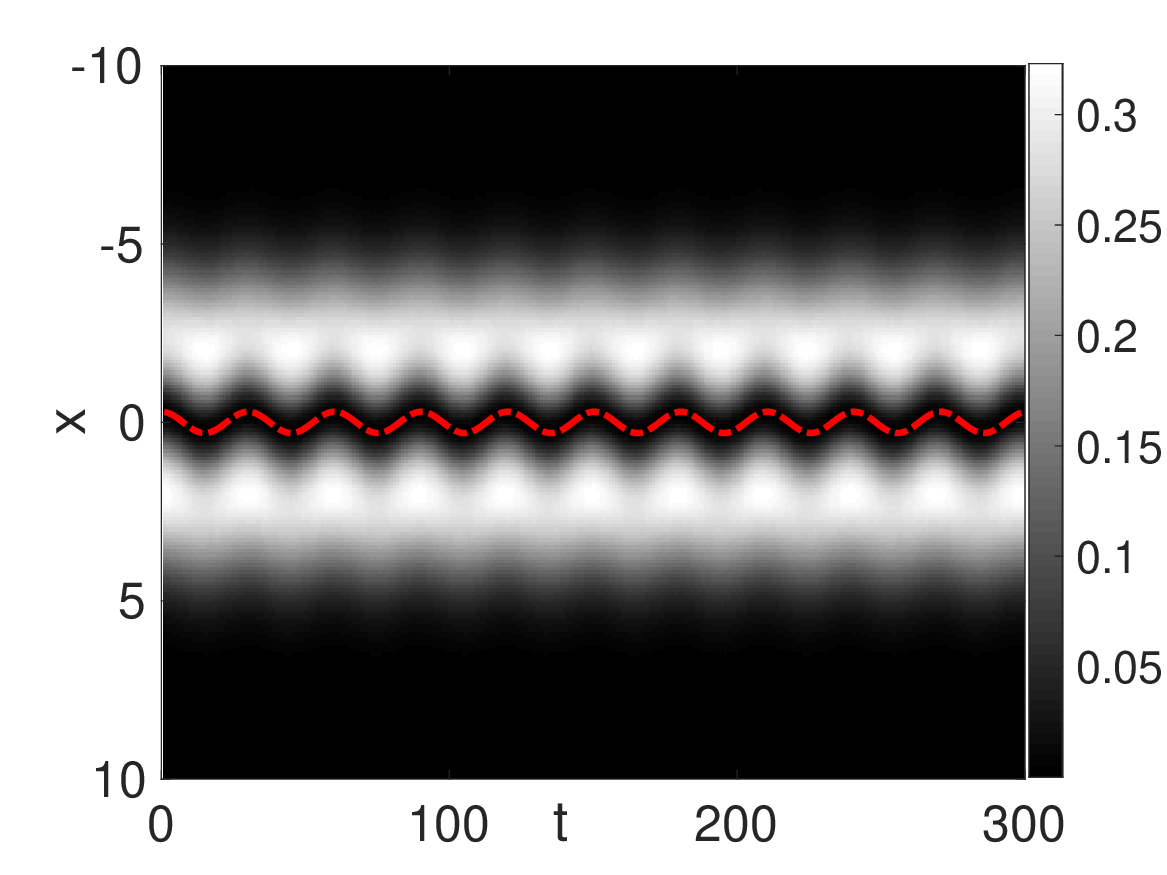}
    \caption{Contour plot of the dynamical space (x)-time (t) evolution of
    a single dark soliton. The colorbar indicates the modulus $|\Psi|$ of
    the wavefunction. The initial condition contains a dark soliton
    perturbed by the (anomalous) eigenmode associated with the dark
    soliton in-trap oscillation. As expected, this leads to a soliton
    oscillation with the frequency predicted by the BdG analysis
    of Eqs.~(\ref{bdg}), namely
    for this case of $\beta=\Omega=0.1$, $\omega=0.2089$. The dashed (red) 
    line shows a simple cosinusoidal curve with this frequency, illustrating
    excellent agreement with the BdG prediction.}
    \label{dyn_ds1}
\end{figure}

Typical examples in the context of the long-range interactions problem 
for the case of the single dark soliton 
are depicted in Figs.~\ref{ex_stab_ds1} and \ref{dyn_ds1}. The
former, characterizes the existence and stability of the 
\panos{numerically obtained solution from Eqs.~(\ref{adc2})-(\ref{adc2b}) 
---with dashed line representing the local and solid the nonlocal
case---} and the latter
encompasses its typical dynamics. The profiles of the top panel
of Fig.~\ref{ex_stab_ds1} are associated with $\beta=0$ (i.e., the
purely local case) and $\beta=\Omega=0.1$, i.e., the case where 
both local and nonlocal interactions are present. Notice that the 
chemical potential used in the top two rows 
is $\mu=1 \gg \Omega$, so we are
close (but not ``at'') the Thomas-Fermi regime. Indeed the former
case of $\beta=0$ resembles closely a $\tanh$-shaped (stationary, 
i.e., bearing vanishing speed) dark 
soliton embedded into (i.e., multiplied by) 
a background of the TF profile
$n_{TF}=\max(0,\mu-V(x))$. On the other hand, in the presence
of nonlocality, we can see that both the local and nonlocal terms
contribute to the profile of the waveform, which maintains its
antisymmetry and the associated $\pi$ phase shift, yet it ``shrinks'' in amplitude, as well as in width.

The second row of Fig.~\ref{ex_stab_ds1} depicts the results of the BdG
analysis, i.e., the lowest modes thereof, including the $\omega=0$
mode due to the U(1) (phase) invariance of the model. The left panel
corresponds to $\beta=0$, a case that is well-studied~\cite{djf,siambook},
while the right panel illustrates the modification of the relevant frequencies,
upon inclusion of the nonlocality. It is important to highlight first that
the single-soliton state retains its spectral stability {\it throughout}
our continuation between $\beta=0$ and $\beta=\Omega$ that we
have considered herein. This suggests that,
in the presence of nonlocality, the solitary waves remain dynamically robust.
In the case of $\beta=0$, it is known that in addition to the lowest modes
of $\omega=\Omega$ (the so-called dipole frequency) and $\omega=\sqrt{3} 
\Omega$ --- and the rest of the 1D modes of 
$\omega=\left(\sqrt{m (m+1)/2}\right) \Omega$
--- there exists a negative energy (so-called anomalous) mode 
at $\omega=\Omega/\sqrt{2}$
(this prediction originally made in~\cite{busch} is valid at the TF limit), 
as summarized in the reviews of~\cite{djf,siambook}
and observed in the experiments of~\cite{becker,markus1,markus2}.  This mode
indicates the excited nature of the dark soliton state. Importantly,
the right panel illustrates the effect of the nonlocal nonlinearity on all of these
modes. Indeed, we find that all the modes are significantly {\it upshifted}, including the anomalous one, except for the
dipole mode that stays unchanged, being associated with an invariance. 
The (upshifted) anomalous mode is intimately
related to the oscillations of the single dark soliton inside
the trap, while the rest of the modes are associated with the background
intrinsic oscillation modes of the entire boson cloud.  Hence,
we conclude that the shrinkage of the condensate cloud is
accompanied by a substantially shorter-period oscillation of
the dark soliton in this nonlocal setting.

In trying to further capture this mode of in-trap oscillation of the dark
soliton, we will leverage the methodology of~\cite{konotop}
(see also~\cite{AstrakharchikPitaevskii2013} for a generalization
to the Lieb-Liniger setting of a Bose gas with $\delta$-function
repulsive interactions). In accordance
with that, in the TF limit, the energy of a dark soliton moving against
the backdrop of a spatially dependent background density is an adiabatic
invariant in the form:
\begin{eqnarray}
E_{ds}=\frac{4}{3} \left(n(x_0) - \dot{x}_0^2 \right)^{3/2},
\label{kk_prl}
\end{eqnarray}
where $x_0$ is the soliton center (and, accordingly, $\dot{x}_0$ is the soliton velocity). 
Upon multiplication by the constant factor (of $3/4$), raising to the
power (of $2/3$) and differentiating Eq.~(\ref{kk_prl}), one obtains
an effective equation for the motion of the dark soliton which can
be combined with Eq.~(\ref{eqn_tf1}) as follows:
\begin{eqnarray}
\ddot{x}_0 = \frac{1}{2} \frac{dn}{dx}\Big|_{x=x_0} \approx - A \sqrt{\frac{\beta}{2}} 
\sin\left( \sqrt{2 \beta} x_0 \right),
\label{kk_prl2}
\end{eqnarray}
with the latter equation being valid in the TF limit and for
$\beta \neq 0$. For oscillations
of the single dark soliton around the origin, 
a Taylor expansion and a choice of a mode of vibration
$x_0 \sim e^{i \omega t}$
yields
an oscillatory motion with a frequency $\omega \approx \sqrt{A \beta}$.
It is relevant to also note here that the frequency $\omega$
depends on $\Omega$ implicitly via the dependence of $A$ on $\Omega$
as per our  earlier discussion.
It is this vibrational mode that we test in the bottom two rows of
Fig.~\ref{ex_stab_ds1} for $\mu=10$ (again for $\Omega=0.1$). We find that this prediction
enables us to capture the relevant oscillation mode not only
in the local interactions case of $\beta=0$ (bottom left panel),
but also adequately in the nonlocal case of $\beta=0.1$ (bottom right panel).
In the latter, the numerical eigenfrequency of the anomalous mode
is found to be $\omega/\Omega=7.3$, while the corresponding theoretical
prediction is $\omega/\Omega=8.02$, arising since $\omega=\sqrt{6.4371 \times 0.1}$, for a relative error of less than $10$\%,
which is quite reasonable given the approximate nature of the calculation,
the narrow nature of the nonlocal waveform in that limit and the 
comparatively wide nature of the dark soliton in this setting.


It is this anomalous mode that we seek to excite in Fig.~\ref{dyn_ds1}.
In particular, we add to the (numerically) exact stationary solution of
Fig.~\ref{ex_stab_ds1} for $\beta=\Omega=0.1$ for $\mu=1$ a significant perturbation 
along the relevant eigendirection. Naturally,
this mode initially displaces the dark soliton, which, in turn, 
executes highly ordered oscillations inside the trap; indeed,
notice that our perturbation is strong enough that it also mildly
excites the ``background'' of the dark soliton. Nevertheless, this
does not affect the accuracy of the result of the linearized prediction
when compared with the direct numerical simulation. Indeed, the relevant
eigenfrequency is $\approx 2.089 \Omega=0.2089$ and it is that
frequency that we very accurately find manifested in the relevant
oscillations of the dark soliton center. A simple cosinusoidal motion
with this frequency is overlaid for definiteness in the corresponding
dynamics of Fig.~\ref{dyn_ds1} with a dashed (red) line as a guide to the eye.

\begin{figure}[tbp]
    \includegraphics[width=0.4\textwidth]{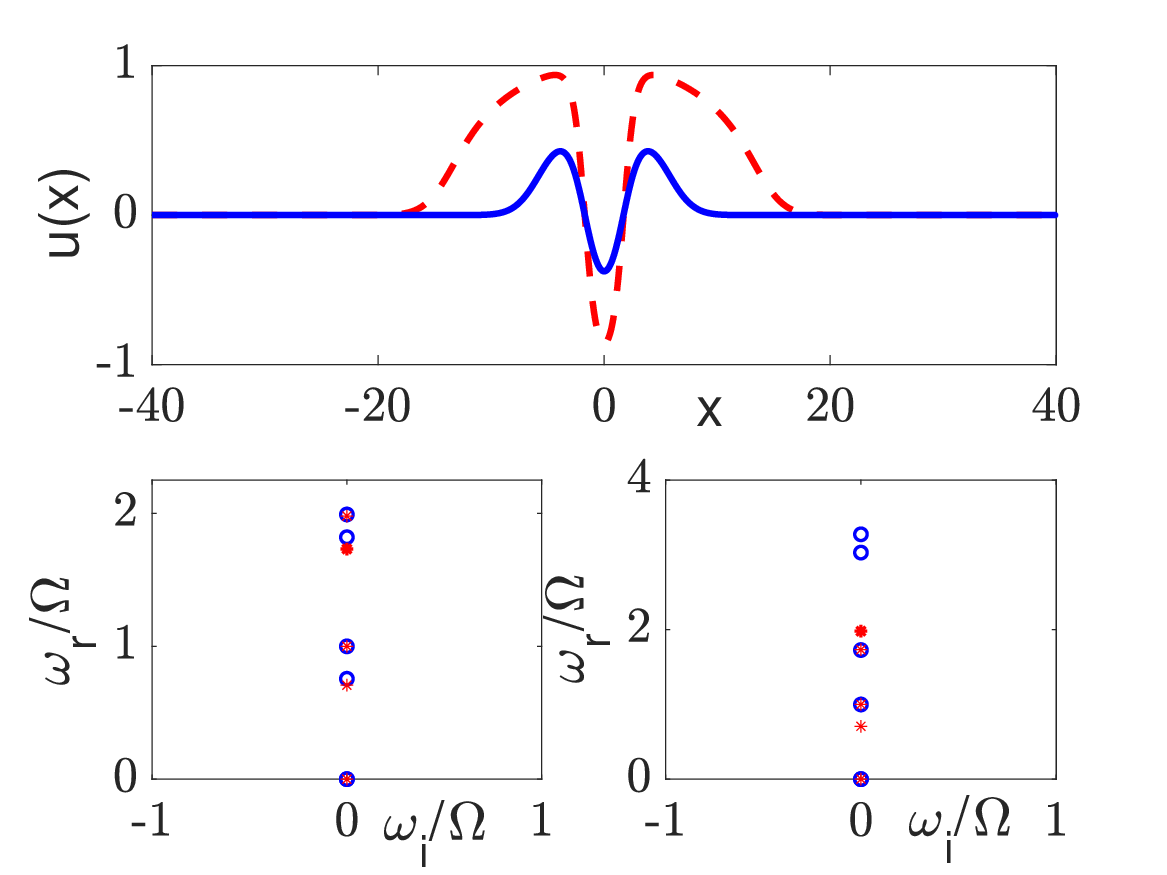}
    \caption{Comparison of the density profile of  two dark solitons  with  and without long-range interactions, represented by a solid and dashed line, respectively. The corresponding collective frequencies are shown as in Fig. \ref{ex_stab_ds1}.  Here, only the case of
    $\mu=1$ is shown.}
    \label{ex_stab_ds3}
\end{figure}

In a similar vein, we can explore the configuration involving two
dark solitons
(again numerically obtained from solving Eqs.~(\ref{adc2})-(\ref{adc2b})), as shown in Fig.~\ref{ex_stab_ds3}.
Here, there exist two anomalous modes, associated with negative
energy, as discussed in~\cite{markus2,djf,siambook}, already at the local
limit of $\beta=0$. One of these modes (the lowest nonzero
frequency of the BdG spectrum) corresponds to the in-phase
oscillation of the two dark solitons with the same
frequency as that of a single soliton, while the other one
corresponds to the out-of-phase motion that has been experimentally
observed~\cite{markus2,markus1}.
Indeed, in the $\beta=0$ limit, both the relative positions
of the solitary waves and the vibration mode frequencies
can be predicted. In particular, according to the prediction
of~\cite{markus2}, the solitary wave positions are found to be
$x_1=-x_2=(1/4) w(64/\Omega^2)$, where $w$ is the Lambert $w$-function,
which is defined as the inverse of $\eta(w)=w e^{w}$. This
prediction yields $x_1=-x_2=1.7103$ for the choice of $\Omega=0.1$,
while numerically we find $x_1=-x_2=1.7198$ 
(from the location of the zero-crossings of the numerical
solution, signaling the soliton positions)
in very good agreement
with the theory, confirming that we are close to the TF limit for
the local nonlinearity case.
The corresponding BdG modes are for the in-phase vibration: 
$\omega_1=\Omega/\sqrt{2}$, while for the out-of-phase one: 
$\omega_2=\sqrt{\Omega^2+64 \exp(-4 x_1)}/\sqrt{2}$. Here, for
instance the latter mode is theoretically predicted to have
$\omega_2=0.1980$ and is numerically found to have 
$\omega_2=0.1992$, i.e., nearly at $2 \Omega$.

These BdG modes, analogously to what we had observed in 
the case of a single dark soliton are significantly upshifted
in frequency as $\beta$ increases. For instance, in the case of
$\beta=\Omega=0.1$, we find that the lower in-phase
oscillation is associated with a frequency of 
$\omega_{IP}=0.1728$ (while this frequency was $0.0756$
i.e., close to $\Omega/\sqrt{2}$, indeed well below
the trap frequency $\Omega=0.1$, in the local case of $\beta=0$).
On the other hand, the higher out-of-phase oscillation
is found to be $\omega_{OP}=0.3276$. 
The relevant eigenfrequencies are illustrated
in the BdG analysis of the bottom panels of Fig.~\ref{ex_stab_ds3},
both for the local case of $\beta=0$ (incorporating also
the analytical predictions via red stars, for the anomalous
modes and the asymptotic frequencies of the ground state TF cloud),
and for the nonlocal one of $\beta \neq 0$.

\begin{figure}[tbp]
    \includegraphics[width=0.4\textwidth]{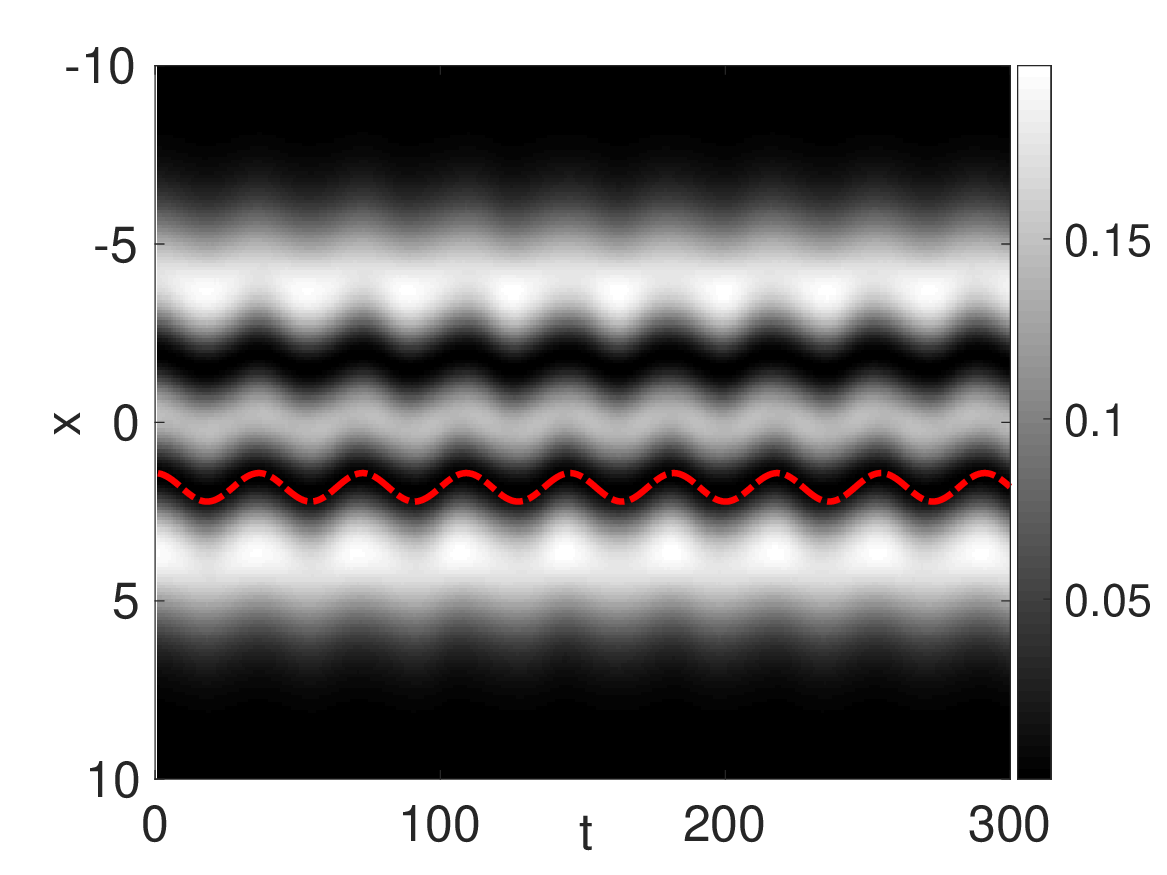}
    \includegraphics[width=0.4\textwidth]{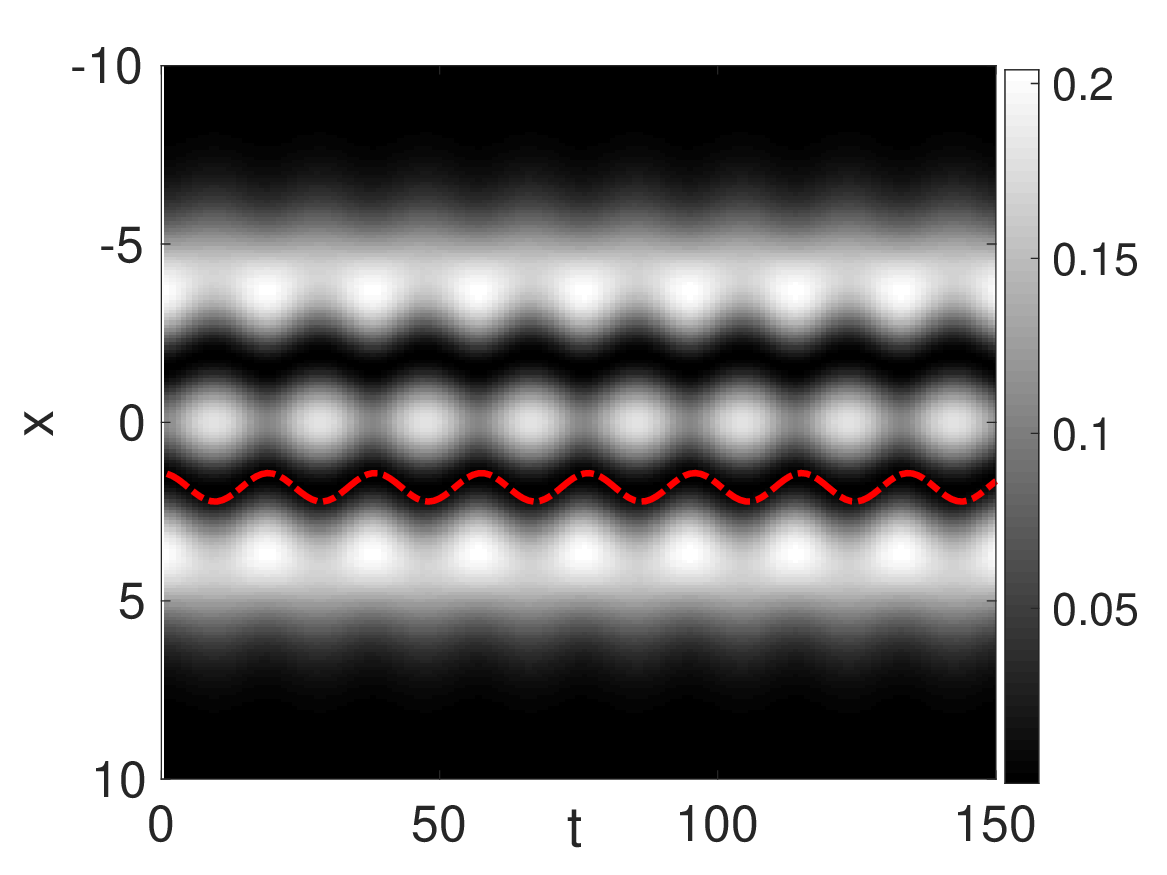}
    \caption{Top panel: similar to Fig.~\ref{dyn_ds1}, but now for the 
    in-phase dynamics
    of a two-soliton state.
The colorbar once again indicates the modulus $|\Psi|$ of
    the wavefunction.
    The state oscillates with a frequency
    $\omega_{IP}=0.1728$ identified in the BdG analysis. Indeed,
    as a guide to the eye for the motion of one of the solitons,
    the (dashed) curve which is cosinusoidal with
    the same frequency is also shown to 
    illustrate the accuracy of the relevant frequency of vibration.
    Bottom panel: same as the top panel but now for the
    out-of-phase oscillation of the two dark solitons with 
    $\omega_{OP}=0.3276$. Once again, the dashed (red) curve 
    represents a cosinusoidal oscillation that
    is superposed as a guide to the eye.}
    \label{dyn_ds2}
\end{figure}

To explore the dynamics associated with these solitonic
(negative energy) eigenmodes in the nonlocal case, we have
perturbed the corresponding eigendirections in the dynamics 
of Eq.~(\ref{eqn1}). Indeed, in each
one of the cases presented in Fig.~\ref{dyn_ds2}, we
observe a vibration with the corresponding eigenmode.
The top panel involves initialization of the 
model with the two-soliton solution, perturbed by the 
in-phase eigenvector of the BdG analysis. Accordingly,
we can observe that the two solitons execute 
robust oscillations with the corresponding in-phase
frequency ($\omega_{\rm IP}=0.1728$). On the other hand, a similar initialization
is performed in the bottom panel, with the only
difference that now we have ``kicked'' the two-soliton
configuration along the eigendirection of the out-of-phase
vibration between the coherent structures. As a result,
in the latter case, we observe a vibration with 
the relevant out-of-phase frequency ($\omega_{\rm OP}=0.3276$).
This pattern can naturally be extended to arbitrary numbers
of dark solitons, with the number of negative energy
modes being equal to the number of dark soliton states 
within the configuration, reflecting the
corresponding excited nature of the state at hand~\cite{siambook}.

\section{Discussion and conclusions}

In the present work, we have explored some aspects of the nonlinear physics of the 
long-range Lieb-Liniger model. The latter constitutes  a deformation of the one-dimensional 
Bose gas with contact interactions (canonical Lieb-Liniger model) resulting  in the case of  
embedding in a harmonic trap, which gives rise to a long-range two-body interaction  term \cite{Beau20,delcampo20}. 
Earlier, in this setting, it was found that ---for attractive local interactions--- the ground state of this model is a trapped bright quantum soliton of the McGuire form. 
Here, we have considered the case of repulsive local interactions, and  
investigated the existence of dark soliton solutions in the mean-field regime, 
that is described by a nonlinear Schr{\"o}dinger (NLS)
equation, 
incorporating the effect of long-range interactions.
To this end, 
upon identifying the relevant density profiles via a fixed-point iteration,
we have performed a Bogolyubov-de Gennes  spectral analysis of single 
and multiple dark soliton states, identifying the characteristic frequency 
describing the evolution of their density profile.
Subsequently, we have confirmed the results of the BdG analysis,
through nonlinear model simulations, confirming the vibrational
modes identified (including the two anomalous ones, describing in- and out-of-phase
oscillations of the two dark solitons).

Our results motivate the quest for many-body quantum soliton wavefunctions 
exhibiting an analogous behavior.
Moreover, there are numerous concrete explorations that the present work
motivates from a nonlinear dynamical perspective. 
More specifically, a natural question is whether the 
asymptotic frequencies of the ground state BdG analysis can be
obtained for the nonlocal case 
in analogy with what is known for the local 
one~\cite{pethick, stringari}. Another is whether the
particle approach developed for a single soliton can
be generalized to multiple solitons as in the work of~\cite{markus2}.
Furthermore, the present analysis has been limited
so far to  a one-dimensional setting. Yet, it would
be particularly interesting and relevant to explore the
extensions to higher dimensional structures and, in particular,
to vortical density profiles~\cite{fetter,siambook}.  

\acknowledgements
It is a pleasure to acknowledge stimulating discussions with Gregory E. Astrakharchik.
This material is based upon work supported by the US National Science Foundation under Grant No. PHY-2110030 (P.G.K.).

\bibliography{DS_LRIbib}

\begin{thebibliography}{38}%
\makeatletter
\providecommand \@ifxundefined [1]{%
 \@ifx{#1\undefined}
}%
\providecommand \@ifnum [1]{%
 \ifnum #1\expandafter \@firstoftwo
 \else \expandafter \@secondoftwo
 \fi
}%
\providecommand \@ifx [1]{%
 \ifx #1\expandafter \@firstoftwo
 \else \expandafter \@secondoftwo
 \fi
}%
\providecommand \natexlab [1]{#1}%
\providecommand \enquote  [1]{``#1''}%
\providecommand \bibnamefont  [1]{#1}%
\providecommand \bibfnamefont [1]{#1}%
\providecommand \citenamefont [1]{#1}%
\providecommand \href@noop [0]{\@secondoftwo}%
\providecommand \href [0]{\begingroup \@sanitize@url \@href}%
\providecommand \@href[1]{\@@startlink{#1}\@@href}%
\providecommand \@@href[1]{\endgroup#1\@@endlink}%
\providecommand \@sanitize@url [0]{\catcode `\\12\catcode `\$12\catcode
  `\&12\catcode `\#12\catcode `\^12\catcode `\_12\catcode `\%12\relax}%
\providecommand \@@startlink[1]{}%
\providecommand \@@endlink[0]{}%
\providecommand \url  [0]{\begingroup\@sanitize@url \@url }%
\providecommand \@url [1]{\endgroup\@href {#1}{\urlprefix }}%
\providecommand \urlprefix  [0]{URL }%
\providecommand \Eprint [0]{\href }%
\providecommand \doibase [0]{http://dx.doi.org/}%
\providecommand \selectlanguage [0]{\@gobble}%
\providecommand \bibinfo  [0]{\@secondoftwo}%
\providecommand \bibfield  [0]{\@secondoftwo}%
\providecommand \translation [1]{[#1]}%
\providecommand \BibitemOpen [0]{}%
\providecommand \bibitemStop [0]{}%
\providecommand \bibitemNoStop [0]{.\EOS\space}%
\providecommand \EOS [0]{\spacefactor3000\relax}%
\providecommand \BibitemShut  [1]{\csname bibitem#1\endcsname}%
\let\auto@bib@innerbib\@empty
\bibitem [{\citenamefont {Lieb}\ and\ \citenamefont {Liniger}(1963)}]{LL63}%
  \BibitemOpen
  \bibfield  {author} {\bibinfo {author} {\bibfnamefont {E.~H.}\ \bibnamefont
  {Lieb}}\ and\ \bibinfo {author} {\bibfnamefont {W.}~\bibnamefont {Liniger}},\
  }\href {\doibase 10.1103/PhysRev.130.1605} {\bibfield  {journal} {\bibinfo
  {journal} {Phys. Rev.}\ }\textbf {\bibinfo {volume} {130}},\ \bibinfo {pages}
  {1605} (\bibinfo {year} {1963})}\BibitemShut {NoStop}%
\bibitem [{\citenamefont {Lieb}(1963)}]{Lieb63}%
  \BibitemOpen
  \bibfield  {author} {\bibinfo {author} {\bibfnamefont {E.~H.}\ \bibnamefont
  {Lieb}},\ }\href {\doibase 10.1103/PhysRev.130.1616} {\bibfield  {journal}
  {\bibinfo  {journal} {Phys. Rev.}\ }\textbf {\bibinfo {volume} {130}},\
  \bibinfo {pages} {1616} (\bibinfo {year} {1963})}\BibitemShut {NoStop}%
\bibitem [{\citenamefont {Sutherland}(2004)}]{Sutherland04}%
  \BibitemOpen
  \bibfield  {author} {\bibinfo {author} {\bibfnamefont {B.}~\bibnamefont
  {Sutherland}},\ }\href@noop {} {\emph {\bibinfo {title} {Beautiful Models: 70
  Years of Exactly Solved Quantum Many-body Problems}}}\ (\bibinfo  {publisher}
  {World Scientific, Singapore},\ \bibinfo {year} {2004})\BibitemShut {NoStop}%
\bibitem [{\citenamefont {Korepin}\ \emph {et~al.}(1997)\citenamefont
  {Korepin}, \citenamefont {Bogoliubov},\ and\ \citenamefont
  {Izergin}}]{KBI97}%
  \BibitemOpen
  \bibfield  {author} {\bibinfo {author} {\bibfnamefont {V.~E.}\ \bibnamefont
  {Korepin}}, \bibinfo {author} {\bibfnamefont {N.~M.}\ \bibnamefont
  {Bogoliubov}}, \ and\ \bibinfo {author} {\bibfnamefont {A.~G.}\ \bibnamefont
  {Izergin}},\ }\href@noop {} {\emph {\bibinfo {title} {Quantum Inverse
  Scattering Method and Correlation Functions}}}\ (\bibinfo  {publisher}
  {Cambridge, Cambridge},\ \bibinfo {year} {1997})\BibitemShut {NoStop}%
\bibitem [{\citenamefont {Takahashi}(1999)}]{Takahashi99}%
  \BibitemOpen
  \bibfield  {author} {\bibinfo {author} {\bibfnamefont {M.}~\bibnamefont
  {Takahashi}},\ }\href@noop {} {\emph {\bibinfo {title} {Thermodynamics of
  One-Dimensional Solvable Models}}}\ (\bibinfo  {publisher} {Cambridge,
  Cambridge},\ \bibinfo {year} {1999})\BibitemShut {NoStop}%
\bibitem [{\citenamefont {Gaudin}(2014)}]{Gaudin14}%
  \BibitemOpen
  \bibfield  {author} {\bibinfo {author} {\bibfnamefont {M.}~\bibnamefont
  {Gaudin}},\ }\href {\doibase 10.1017/CBO9781107053885} {\emph {\bibinfo
  {title} {The Bethe Wavefunction}}},\ edited by\ \bibinfo {editor}
  {\bibfnamefont {J.-S.}\ \bibnamefont {Caux}}\ (\bibinfo  {publisher}
  {Cambridge University Press},\ \bibinfo {year} {2014})\BibitemShut {NoStop}%
\bibitem [{\citenamefont {Olshanii}(1998)}]{Olshanii98}%
  \BibitemOpen
  \bibfield  {author} {\bibinfo {author} {\bibfnamefont {M.}~\bibnamefont
  {Olshanii}},\ }\href {\doibase 10.1103/PhysRevLett.81.938} {\bibfield
  {journal} {\bibinfo  {journal} {Phys. Rev. Lett.}\ }\textbf {\bibinfo
  {volume} {81}},\ \bibinfo {pages} {938} (\bibinfo {year} {1998})}\BibitemShut
  {NoStop}%
\bibitem [{\citenamefont {Cazalilla}\ \emph {et~al.}(2011)\citenamefont
  {Cazalilla}, \citenamefont {Citro}, \citenamefont {Giamarchi}, \citenamefont
  {Orignac},\ and\ \citenamefont {Rigol}}]{Cazalilla11}%
  \BibitemOpen
  \bibfield  {author} {\bibinfo {author} {\bibfnamefont {M.~A.}\ \bibnamefont
  {Cazalilla}}, \bibinfo {author} {\bibfnamefont {R.}~\bibnamefont {Citro}},
  \bibinfo {author} {\bibfnamefont {T.}~\bibnamefont {Giamarchi}}, \bibinfo
  {author} {\bibfnamefont {E.}~\bibnamefont {Orignac}}, \ and\ \bibinfo
  {author} {\bibfnamefont {M.}~\bibnamefont {Rigol}},\ }\href {\doibase
  10.1103/RevModPhys.83.1405} {\bibfield  {journal} {\bibinfo  {journal} {Rev.
  Mod. Phys.}\ }\textbf {\bibinfo {volume} {83}},\ \bibinfo {pages} {1405}
  (\bibinfo {year} {2011})}\BibitemShut {NoStop}%
\bibitem [{\citenamefont {Beau}\ \emph {et~al.}(2020)\citenamefont {Beau},
  \citenamefont {Pittman}, \citenamefont {Astrakharchik},\ and\ \citenamefont
  {del Campo}}]{Beau20}%
  \BibitemOpen
  \bibfield  {author} {\bibinfo {author} {\bibfnamefont {M.}~\bibnamefont
  {Beau}}, \bibinfo {author} {\bibfnamefont {S.~M.}\ \bibnamefont {Pittman}},
  \bibinfo {author} {\bibfnamefont {G.~E.}\ \bibnamefont {Astrakharchik}}, \
  and\ \bibinfo {author} {\bibfnamefont {A.}~\bibnamefont {del Campo}},\ }\href
  {\doibase 10.1103/PhysRevLett.125.220602} {\bibfield  {journal} {\bibinfo
  {journal} {Phys. Rev. Lett.}\ }\textbf {\bibinfo {volume} {125}},\ \bibinfo
  {pages} {220602} (\bibinfo {year} {2020})}\BibitemShut {NoStop}%
\bibitem [{\citenamefont {del Campo}(2020)}]{delcampo20}%
  \BibitemOpen
  \bibfield  {author} {\bibinfo {author} {\bibfnamefont {A.}~\bibnamefont {del
  Campo}},\ }\href {\doibase 10.1103/PhysRevResearch.2.043114} {\bibfield
  {journal} {\bibinfo  {journal} {Phys. Rev. Research}\ }\textbf {\bibinfo
  {volume} {2}},\ \bibinfo {pages} {043114} (\bibinfo {year}
  {2020})}\BibitemShut {NoStop}%
\bibitem [{\citenamefont {Lieb}\ \emph {et~al.}(2018)\citenamefont {Lieb},
  \citenamefont {Rougerie},\ and\ \citenamefont {Yngvason}}]{Lieb2018}%
  \BibitemOpen
  \bibfield  {author} {\bibinfo {author} {\bibfnamefont {E.~H.}\ \bibnamefont
  {Lieb}}, \bibinfo {author} {\bibfnamefont {N.}~\bibnamefont {Rougerie}}, \
  and\ \bibinfo {author} {\bibfnamefont {J.}~\bibnamefont {Yngvason}},\ }\href
  {\doibase 10.1007/s10955-018-2082-1} {\bibfield  {journal} {\bibinfo
  {journal} {Journal of Statistical Physics}\ }\textbf {\bibinfo {volume}
  {172}},\ \bibinfo {pages} {544} (\bibinfo {year} {2018})}\BibitemShut
  {NoStop}%
\bibitem [{\citenamefont {Schive}\ \emph {et~al.}(2014)\citenamefont {Schive},
  \citenamefont {Chiueh},\ and\ \citenamefont {Broadhurst}}]{Broadhurst14}%
  \BibitemOpen
  \bibfield  {author} {\bibinfo {author} {\bibfnamefont {H.-Y.}\ \bibnamefont
  {Schive}}, \bibinfo {author} {\bibfnamefont {T.}~\bibnamefont {Chiueh}}, \
  and\ \bibinfo {author} {\bibfnamefont {T.}~\bibnamefont {Broadhurst}},\
  }\href {\doibase 10.1038/nphys2996} {\bibfield  {journal} {\bibinfo
  {journal} {Nature Physics}\ }\textbf {\bibinfo {volume} {10}},\ \bibinfo
  {pages} {496} (\bibinfo {year} {2014})}\BibitemShut {NoStop}%
\bibitem [{\citenamefont {Broadhurst}\ \emph {et~al.}(2020)\citenamefont
  {Broadhurst}, \citenamefont {De~Martino}, \citenamefont {Luu}, \citenamefont
  {Smoot},\ and\ \citenamefont {Tye}}]{Broadhurst20}%
  \BibitemOpen
  \bibfield  {author} {\bibinfo {author} {\bibfnamefont {T.}~\bibnamefont
  {Broadhurst}}, \bibinfo {author} {\bibfnamefont {I.}~\bibnamefont
  {De~Martino}}, \bibinfo {author} {\bibfnamefont {H.~N.}\ \bibnamefont {Luu}},
  \bibinfo {author} {\bibfnamefont {G.~F.}\ \bibnamefont {Smoot}}, \ and\
  \bibinfo {author} {\bibfnamefont {S.-H.~H.}\ \bibnamefont {Tye}},\ }\href
  {\doibase 10.1103/PhysRevD.101.083012} {\bibfield  {journal} {\bibinfo
  {journal} {Phys. Rev. D}\ }\textbf {\bibinfo {volume} {101}},\ \bibinfo
  {pages} {083012} (\bibinfo {year} {2020})}\BibitemShut {NoStop}%
\bibitem [{\citenamefont {Beau}\ and\ \citenamefont {del
  Campo}(2021)}]{Beau21}%
  \BibitemOpen
  \bibfield  {author} {\bibinfo {author} {\bibfnamefont {M.}~\bibnamefont
  {Beau}}\ and\ \bibinfo {author} {\bibfnamefont {A.}~\bibnamefont {del
  Campo}},\ }\href@noop {} {\enquote {\bibinfo {title} {Parent hamiltonians of
  jastrow wavefunctions},}\ } (\bibinfo {year} {2021}),\ \Eprint
  {http://arxiv.org/abs/2107.02869} {arXiv:2107.02869 [nlin.SI]} \BibitemShut
  {NoStop}%
\bibitem [{\citenamefont {Kundu}(2009)}]{Kundu09}%
  \BibitemOpen
  \bibfield  {author} {\bibinfo {author} {\bibfnamefont {A.}~\bibnamefont
  {Kundu}},\ }\href {\doibase 10.1103/PhysRevE.79.015601} {\bibfield  {journal}
  {\bibinfo  {journal} {Phys. Rev. E}\ }\textbf {\bibinfo {volume} {79}},\
  \bibinfo {pages} {015601} (\bibinfo {year} {2009})}\BibitemShut {NoStop}%
\bibitem [{\citenamefont {McGuire}(1964)}]{McGuire64}%
  \BibitemOpen
  \bibfield  {author} {\bibinfo {author} {\bibfnamefont {J.~B.}\ \bibnamefont
  {McGuire}},\ }\href {\doibase 10.1063/1.1704156} {\bibfield  {journal}
  {\bibinfo  {journal} {Journal of Mathematical Physics}\ }\textbf {\bibinfo
  {volume} {5}},\ \bibinfo {pages} {622} (\bibinfo {year} {1964})}\BibitemShut
  {NoStop}%
\bibitem [{\citenamefont {Kulish}\ \emph {et~al.}(1976)\citenamefont {Kulish},
  \citenamefont {Manakov},\ and\ \citenamefont {Faddeev}}]{Kulish76}%
  \BibitemOpen
  \bibfield  {author} {\bibinfo {author} {\bibfnamefont {P.~P.}\ \bibnamefont
  {Kulish}}, \bibinfo {author} {\bibfnamefont {S.~V.}\ \bibnamefont {Manakov}},
  \ and\ \bibinfo {author} {\bibfnamefont {L.~D.}\ \bibnamefont {Faddeev}},\
  }\href {\doibase 10.1007/BF01028912} {\bibfield  {journal} {\bibinfo
  {journal} {Theoretical and Mathematical Physics}\ }\textbf {\bibinfo {volume}
  {28}},\ \bibinfo {pages} {615} (\bibinfo {year} {1976})}\BibitemShut
  {NoStop}%
\bibitem [{\citenamefont {Ishikawa}\ and\ \citenamefont
  {Takayama}(1980)}]{Ishikawa80}%
  \BibitemOpen
  \bibfield  {author} {\bibinfo {author} {\bibfnamefont {M.}~\bibnamefont
  {Ishikawa}}\ and\ \bibinfo {author} {\bibfnamefont {H.}~\bibnamefont
  {Takayama}},\ }\href {\doibase 10.1143/JPSJ.49.1242} {\bibfield  {journal}
  {\bibinfo  {journal} {Journal of the Physical Society of Japan}\ }\textbf
  {\bibinfo {volume} {49}},\ \bibinfo {pages} {1242} (\bibinfo {year}
  {1980})}\BibitemShut {NoStop}%
\bibitem [{\citenamefont {Tsuzuki}(1971)}]{Tsuzuki71}%
  \BibitemOpen
  \bibfield  {author} {\bibinfo {author} {\bibfnamefont {T.}~\bibnamefont
  {Tsuzuki}},\ }\href {\doibase 10.1007/BF00628744} {\bibfield  {journal}
  {\bibinfo  {journal} {Journal of Low Temperature Physics}\ }\textbf {\bibinfo
  {volume} {4}},\ \bibinfo {pages} {441} (\bibinfo {year} {1971})}\BibitemShut
  {NoStop}%
\bibitem [{\citenamefont {Sato}\ \emph {et~al.}(2012)\citenamefont {Sato},
  \citenamefont {Kanamoto}, \citenamefont {Kaminishi},\ and\ \citenamefont
  {Deguchi}}]{Deguchi1}%
  \BibitemOpen
  \bibfield  {author} {\bibinfo {author} {\bibfnamefont {J.}~\bibnamefont
  {Sato}}, \bibinfo {author} {\bibfnamefont {R.}~\bibnamefont {Kanamoto}},
  \bibinfo {author} {\bibfnamefont {E.}~\bibnamefont {Kaminishi}}, \ and\
  \bibinfo {author} {\bibfnamefont {T.}~\bibnamefont {Deguchi}},\ }\href
  {\doibase 10.1103/PhysRevLett.108.110401} {\bibfield  {journal} {\bibinfo
  {journal} {Phys. Rev. Lett.}\ }\textbf {\bibinfo {volume} {108}},\ \bibinfo
  {pages} {110401} (\bibinfo {year} {2012})}\BibitemShut {NoStop}%
\bibitem [{\citenamefont {Sato}\ \emph {et~al.}(2016)\citenamefont {Sato},
  \citenamefont {Kanamoto}, \citenamefont {Kaminishi},\ and\ \citenamefont
  {Deguchi}}]{Deguchi2}%
  \BibitemOpen
  \bibfield  {author} {\bibinfo {author} {\bibfnamefont {J.}~\bibnamefont
  {Sato}}, \bibinfo {author} {\bibfnamefont {R.}~\bibnamefont {Kanamoto}},
  \bibinfo {author} {\bibfnamefont {E.}~\bibnamefont {Kaminishi}}, \ and\
  \bibinfo {author} {\bibfnamefont {T.}~\bibnamefont {Deguchi}},\ }\href
  {\doibase 10.1088/1367-2630/18/7/075008} {\bibfield  {journal} {\bibinfo
  {journal} {New Journal of Physics}\ }\textbf {\bibinfo {volume} {18}},\
  \bibinfo {pages} {075008} (\bibinfo {year} {2016})}\BibitemShut {NoStop}%
\bibitem [{\citenamefont {Girardeau}\ and\ \citenamefont
  {Wright}(2000)}]{GW00}%
  \BibitemOpen
  \bibfield  {author} {\bibinfo {author} {\bibfnamefont {M.~D.}\ \bibnamefont
  {Girardeau}}\ and\ \bibinfo {author} {\bibfnamefont {E.~M.}\ \bibnamefont
  {Wright}},\ }\href {\doibase 10.1103/PhysRevLett.84.5691} {\bibfield
  {journal} {\bibinfo  {journal} {Phys. Rev. Lett.}\ }\textbf {\bibinfo
  {volume} {84}},\ \bibinfo {pages} {5691} (\bibinfo {year}
  {2000})}\BibitemShut {NoStop}%
\bibitem [{\citenamefont {Koutsokostas}\ \emph {et~al.}(2020)\citenamefont
  {Koutsokostas}, \citenamefont {Horikis}, \citenamefont {Kevrekidis},\ and\
  \citenamefont {Frantzeskakis}}]{djf2}%
  \BibitemOpen
  \bibfield  {author} {\bibinfo {author} {\bibfnamefont {G.~N.}\ \bibnamefont
  {Koutsokostas}}, \bibinfo {author} {\bibfnamefont {T.~P.}\ \bibnamefont
  {Horikis}}, \bibinfo {author} {\bibfnamefont {P.~G.}\ \bibnamefont
  {Kevrekidis}}, \ and\ \bibinfo {author} {\bibfnamefont {D.~J.}\ \bibnamefont
  {Frantzeskakis}},\ }\href@noop {} {\enquote {\bibinfo {title} {Universal
  reductions and solitary waves of weakly nonlocal defocusing nonlinear
  schr\"odinger equations},}\ } (\bibinfo {year} {2020}),\ \Eprint
  {http://arxiv.org/abs/2011.09651} {arXiv:2011.09651 [nlin.PS]} \BibitemShut
  {NoStop}%
\bibitem [{\citenamefont {Konotop}\ and\ \citenamefont
  {Pitaevskii}(2004)}]{konotop}%
  \BibitemOpen
  \bibfield  {author} {\bibinfo {author} {\bibfnamefont {V.~V.}\ \bibnamefont
  {Konotop}}\ and\ \bibinfo {author} {\bibfnamefont {L.}~\bibnamefont
  {Pitaevskii}},\ }\href {\doibase 10.1103/PhysRevLett.93.240403} {\bibfield
  {journal} {\bibinfo  {journal} {Phys. Rev. Lett.}\ }\textbf {\bibinfo
  {volume} {93}},\ \bibinfo {pages} {240403} (\bibinfo {year}
  {2004})}\BibitemShut {NoStop}%
\bibitem [{\citenamefont {Petrov}\ \emph {et~al.}(2000)\citenamefont {Petrov},
  \citenamefont {Shlyapnikov},\ and\ \citenamefont {Walraven}}]{Petrov00}%
  \BibitemOpen
  \bibfield  {author} {\bibinfo {author} {\bibfnamefont {D.~S.}\ \bibnamefont
  {Petrov}}, \bibinfo {author} {\bibfnamefont {G.~V.}\ \bibnamefont
  {Shlyapnikov}}, \ and\ \bibinfo {author} {\bibfnamefont {J.~T.~M.}\
  \bibnamefont {Walraven}},\ }\href {\doibase 10.1103/PhysRevLett.85.3745}
  {\bibfield  {journal} {\bibinfo  {journal} {Phys. Rev. Lett.}\ }\textbf
  {\bibinfo {volume} {85}},\ \bibinfo {pages} {3745} (\bibinfo {year}
  {2000})}\BibitemShut {NoStop}%
\bibitem [{\citenamefont {Frantzeskakis}(2010)}]{djf}%
  \BibitemOpen
  \bibfield  {author} {\bibinfo {author} {\bibfnamefont {D.~J.}\ \bibnamefont
  {Frantzeskakis}},\ }\href {\doibase 10.1088/1751-8113/43/21/213001}
  {\bibfield  {journal} {\bibinfo  {journal} {Journal of Physics A:
  Mathematical and Theoretical}\ }\textbf {\bibinfo {volume} {43}},\ \bibinfo
  {pages} {213001} (\bibinfo {year} {2010})}\BibitemShut {NoStop}%
\bibitem [{\citenamefont {Kevrekidis}\ \emph {et~al.}(2015)\citenamefont
  {Kevrekidis}, \citenamefont {Frantzeskakis},\ and\ \citenamefont
  {Carretero-Gonz\'alez}}]{siambook}%
  \BibitemOpen
  \bibfield  {author} {\bibinfo {author} {\bibfnamefont {P.~G.}\ \bibnamefont
  {Kevrekidis}}, \bibinfo {author} {\bibfnamefont {D.~J.}\ \bibnamefont
  {Frantzeskakis}}, \ and\ \bibinfo {author} {\bibfnamefont {R.}~\bibnamefont
  {Carretero-Gonz\'alez}},\ }\href@noop {} {\emph {\bibinfo {title} {The
  Defocusing Nonlinear Schr{\"o}dinger Equation}}}\ (\bibinfo  {publisher}
  {SIAM, Philadelphia},\ \bibinfo {year} {2015})\BibitemShut {NoStop}%
\bibitem [{\citenamefont {Pethick}\ and\ \citenamefont
  {Smith}(2002)}]{pethick}%
  \BibitemOpen
  \bibfield  {author} {\bibinfo {author} {\bibfnamefont {C.~J.}\ \bibnamefont
  {Pethick}}\ and\ \bibinfo {author} {\bibfnamefont {H.}~\bibnamefont
  {Smith}},\ }\href@noop {} {\emph {\bibinfo {title} {Bose-Einstein
  condensation in dilute gases}}}\ (\bibinfo  {publisher} {Cambridge,
  Cambridge},\ \bibinfo {year} {2002})\BibitemShut {NoStop}%
\bibitem [{\citenamefont {Pitaevskii}\ and\ \citenamefont
  {Stringari}(2003)}]{stringari}%
  \BibitemOpen
  \bibfield  {author} {\bibinfo {author} {\bibfnamefont {L.~P.}\ \bibnamefont
  {Pitaevskii}}\ and\ \bibinfo {author} {\bibfnamefont {S.}~\bibnamefont
  {Stringari}},\ }\href@noop {} {\emph {\bibinfo {title} {Bose-Einstein
  Condensation}}}\ (\bibinfo  {publisher} {Oxford, Oxford},\ \bibinfo {year}
  {2003})\BibitemShut {NoStop}%
\bibitem [{\citenamefont {Gallo}\ and\ \citenamefont
  {Pelinovsky}(2011)}]{gallo}%
  \BibitemOpen
  \bibfield  {author} {\bibinfo {author} {\bibfnamefont {C.}~\bibnamefont
  {Gallo}}\ and\ \bibinfo {author} {\bibfnamefont {D.~E.}\ \bibnamefont
  {Pelinovsky}},\ }\href@noop {} {\bibfield  {journal} {\bibinfo  {journal}
  {Asympt. Anal.}\ }\textbf {\bibinfo {volume} {73}},\ \bibinfo {pages} {53}
  (\bibinfo {year} {2011})}\BibitemShut {NoStop}%
\bibitem [{\citenamefont {Menotti}\ and\ \citenamefont
  {Stringari}(2002)}]{menotti}%
  \BibitemOpen
  \bibfield  {author} {\bibinfo {author} {\bibfnamefont {C.}~\bibnamefont
  {Menotti}}\ and\ \bibinfo {author} {\bibfnamefont {S.}~\bibnamefont
  {Stringari}},\ }\href {\doibase 10.1103/PhysRevA.66.043610} {\bibfield
  {journal} {\bibinfo  {journal} {Phys. Rev. A}\ }\textbf {\bibinfo {volume}
  {66}},\ \bibinfo {pages} {043610} (\bibinfo {year} {2002})}\BibitemShut
  {NoStop}%
\bibitem [{\citenamefont {De~Rosi}\ and\ \citenamefont
  {Stringari}(2015)}]{derosi}%
  \BibitemOpen
  \bibfield  {author} {\bibinfo {author} {\bibfnamefont {G.}~\bibnamefont
  {De~Rosi}}\ and\ \bibinfo {author} {\bibfnamefont {S.}~\bibnamefont
  {Stringari}},\ }\href {\doibase 10.1103/PhysRevA.92.053617} {\bibfield
  {journal} {\bibinfo  {journal} {Phys. Rev. A}\ }\textbf {\bibinfo {volume}
  {92}},\ \bibinfo {pages} {053617} (\bibinfo {year} {2015})}\BibitemShut
  {NoStop}%
\bibitem [{\citenamefont {Busch}\ and\ \citenamefont {Anglin}(2000)}]{busch}%
  \BibitemOpen
  \bibfield  {author} {\bibinfo {author} {\bibfnamefont {T.}~\bibnamefont
  {Busch}}\ and\ \bibinfo {author} {\bibfnamefont {J.~R.}\ \bibnamefont
  {Anglin}},\ }\href {\doibase 10.1103/PhysRevLett.84.2298} {\bibfield
  {journal} {\bibinfo  {journal} {Phys. Rev. Lett.}\ }\textbf {\bibinfo
  {volume} {84}},\ \bibinfo {pages} {2298} (\bibinfo {year}
  {2000})}\BibitemShut {NoStop}%
\bibitem [{\citenamefont {Becker}\ \emph {et~al.}(2008)\citenamefont {Becker},
  \citenamefont {Stellmer}, \citenamefont {Soltan-Panahi}, \citenamefont
  {D{\"o}rscher}, \citenamefont {Baumert}, \citenamefont {Richter},
  \citenamefont {Kronj{\"a}ger}, \citenamefont {Bongs},\ and\ \citenamefont
  {Sengstock}}]{becker}%
  \BibitemOpen
  \bibfield  {author} {\bibinfo {author} {\bibfnamefont {C.}~\bibnamefont
  {Becker}}, \bibinfo {author} {\bibfnamefont {S.}~\bibnamefont {Stellmer}},
  \bibinfo {author} {\bibfnamefont {P.}~\bibnamefont {Soltan-Panahi}}, \bibinfo
  {author} {\bibfnamefont {S.}~\bibnamefont {D{\"o}rscher}}, \bibinfo {author}
  {\bibfnamefont {M.}~\bibnamefont {Baumert}}, \bibinfo {author} {\bibfnamefont
  {E.-M.}\ \bibnamefont {Richter}}, \bibinfo {author} {\bibfnamefont
  {J.}~\bibnamefont {Kronj{\"a}ger}}, \bibinfo {author} {\bibfnamefont
  {K.}~\bibnamefont {Bongs}}, \ and\ \bibinfo {author} {\bibfnamefont
  {K.}~\bibnamefont {Sengstock}},\ }\href@noop {} {\bibfield  {journal}
  {\bibinfo  {journal} {Nature Physics}\ }\textbf {\bibinfo {volume} {4}},\
  \bibinfo {pages} {496} (\bibinfo {year} {2008})}\BibitemShut {NoStop}%
\bibitem [{\citenamefont {Weller}\ \emph {et~al.}(2008)\citenamefont {Weller},
  \citenamefont {Ronzheimer}, \citenamefont {Gross}, \citenamefont {Esteve},
  \citenamefont {Oberthaler}, \citenamefont {Frantzeskakis}, \citenamefont
  {Theocharis},\ and\ \citenamefont {Kevrekidis}}]{markus1}%
  \BibitemOpen
  \bibfield  {author} {\bibinfo {author} {\bibfnamefont {A.}~\bibnamefont
  {Weller}}, \bibinfo {author} {\bibfnamefont {J.~P.}\ \bibnamefont
  {Ronzheimer}}, \bibinfo {author} {\bibfnamefont {C.}~\bibnamefont {Gross}},
  \bibinfo {author} {\bibfnamefont {J.}~\bibnamefont {Esteve}}, \bibinfo
  {author} {\bibfnamefont {M.~K.}\ \bibnamefont {Oberthaler}}, \bibinfo
  {author} {\bibfnamefont {D.~J.}\ \bibnamefont {Frantzeskakis}}, \bibinfo
  {author} {\bibfnamefont {G.}~\bibnamefont {Theocharis}}, \ and\ \bibinfo
  {author} {\bibfnamefont {P.~G.}\ \bibnamefont {Kevrekidis}},\ }\href
  {\doibase 10.1103/PhysRevLett.101.130401} {\bibfield  {journal} {\bibinfo
  {journal} {Phys. Rev. Lett.}\ }\textbf {\bibinfo {volume} {101}},\ \bibinfo
  {pages} {130401} (\bibinfo {year} {2008})}\BibitemShut {NoStop}%
\bibitem [{\citenamefont {Theocharis}\ \emph {et~al.}(2010)\citenamefont
  {Theocharis}, \citenamefont {Weller}, \citenamefont {Ronzheimer},
  \citenamefont {Gross}, \citenamefont {Oberthaler}, \citenamefont
  {Kevrekidis},\ and\ \citenamefont {Frantzeskakis}}]{markus2}%
  \BibitemOpen
  \bibfield  {author} {\bibinfo {author} {\bibfnamefont {G.}~\bibnamefont
  {Theocharis}}, \bibinfo {author} {\bibfnamefont {A.}~\bibnamefont {Weller}},
  \bibinfo {author} {\bibfnamefont {J.~P.}\ \bibnamefont {Ronzheimer}},
  \bibinfo {author} {\bibfnamefont {C.}~\bibnamefont {Gross}}, \bibinfo
  {author} {\bibfnamefont {M.~K.}\ \bibnamefont {Oberthaler}}, \bibinfo
  {author} {\bibfnamefont {P.~G.}\ \bibnamefont {Kevrekidis}}, \ and\ \bibinfo
  {author} {\bibfnamefont {D.~J.}\ \bibnamefont {Frantzeskakis}},\ }\href
  {\doibase 10.1103/PhysRevA.81.063604} {\bibfield  {journal} {\bibinfo
  {journal} {Phys. Rev. A}\ }\textbf {\bibinfo {volume} {81}},\ \bibinfo
  {pages} {063604} (\bibinfo {year} {2010})}\BibitemShut {NoStop}%
\bibitem [{\citenamefont {Astrakharchik}\ and\ \citenamefont
  {Pitaevskii}(2013)}]{AstrakharchikPitaevskii2013}%
  \BibitemOpen
  \bibfield  {author} {\bibinfo {author} {\bibfnamefont {G.~E.}\ \bibnamefont
  {Astrakharchik}}\ and\ \bibinfo {author} {\bibfnamefont {L.~P.}\ \bibnamefont
  {Pitaevskii}},\ }\href {\doibase 10.1209/0295-5075/102/30004} {\bibfield
  {journal} {\bibinfo  {journal} {EPL (Europhysics Letters)}\ }\textbf
  {\bibinfo {volume} {102}},\ \bibinfo {pages} {30004} (\bibinfo {year}
  {2013})}\BibitemShut {NoStop}%
\bibitem [{\citenamefont {Fetter}(2009)}]{fetter}%
  \BibitemOpen
  \bibfield  {author} {\bibinfo {author} {\bibfnamefont {A.~L.}\ \bibnamefont
  {Fetter}},\ }\href {\doibase 10.1103/RevModPhys.81.647} {\bibfield  {journal}
  {\bibinfo  {journal} {Rev. Mod. Phys.}\ }\textbf {\bibinfo {volume} {81}},\
  \bibinfo {pages} {647} (\bibinfo {year} {2009})}\BibitemShut {NoStop}%
\end{thebibliography}%

\end{document}